\documentclass[manuscript]{aastex}

\shorttitle{}
\shortauthors{Sharykin and Kosovichev}

\begin{document}

\title{DYNAMICS OF ELECTRIC CURRENTS, MAGNETIC FIELD TOPOLOGY AND HELIOSEISMIC RESPONSE OF A SOLAR FLARE}

\author{I.N. Sharykin\altaffilmark{1,2} and A.G. Kosovichev\altaffilmark{1,3,4}}
\affil{Big Bear Solar Observatory, New Jersey Institute of Technology,
    Big Bear City, CA 92314, U.S.A}

\altaffiltext{1}{Big Bear Solar Observatory}
\altaffiltext{2}{Space Research Institute (IKI) of the Russian Academy of Science}
\altaffiltext{3}{Stanford University}
\altaffiltext{4}{NASA Ames Research Center}


\begin{abstract}

The solar flare on July 30, 2011 was of a modest X-ray class (M9.3), but it made a strong photospheric impact and produced a ``sunquake,'' observed with the Helioseismic and Magnetic Imager (HMI) on NASA's Solar Dynamics Observatory (SDO). In addition to the helioseismic waves (also observed with the  SDO/AIA instrument), the flare caused a large expanding area of white-light emission and was accompanied by substantial restructuring of  magnetic fields, leading to the rapid formation of a sunspot structure in the flare region. The flare produced  no significant hard X-ray emission and no coronal mass ejection. This indicates that the flare energy release was mostly confined to the lower atmosphere. The absence of significant coronal mass ejection rules out magnetic rope eruption as a mechanism of helioseismic waves. We discuss the connectivity of the flare energy release with the electric currents dynamics and show the potential importance of high-speed plasma flows in the lower solar atmosphere during the flare energy release.

\end{abstract}

\section{INTRODUCTION}

The flare energy release is usually accompanied by restructuring of magnetic field due to magnetic reconnection \citep[e.g.][]{Su2013}, dissipation, and eruptions of magnetized plasma. The restructuring is observed in the form of arcades of magnetic loops with reducing shear, changes of the strength and geometry of photospheric magnetic fields, erupting magnetic structures (CMEs), and so on. Sometimes, the magnetic topology of the flare energy release site can be very complicated due to a complex interaction between magnetic fields rooted in multipolar active regions. Despite the complexity, all flares are accompanied by standard physical processes like particle acceleration, plasma heating up to several MK and plasma motion. The standard flare model \citep{Sturrock1989} assumes that the flare energy release processes are caused by magnetic reconnection in the corona, but the idea of flare initiation in the lower solar atmosphere has recently become popular because of new observational evidences~\citep[e.g.][]{Fletcher2011}. Physically, the lower solar atmosphere significantly differs from the corona. The low atmosphere is characterized by partially ionized plasma, higher plasma $\beta$, turbulent motions and highly pronounced NLTE conditions during solar flares. The work of \citet{Leake2012} is an example showing the complexity of magnetic reconnection in the chromospheric plasma and reveals a fast reconnection regime with the rate independent on the plasma resistivity. Generally, the physical processes in the chromospheric conditions are not well-understood, and will be an important topic of future theoretical and observational studies. The current research is devoted to the observational study of flare processes in the lower solar atmosphere.

In this work, as a case study, we present observations and analysis of a moderate M9.3 class flare, which reveal several significant perturbations in the low atmosphere. The flare generated helioseismic waves, caused a large expanding area of continuum emission, and was accompanied by substantial restructuring of  magnetic fields in the flare region, leading to a rapid formation of a sunspot structure in the flare region. The flare produced no significant hard X-ray emission, no coronal mass ejection, and no II and III type radio bursts. This indicates that the flare energy release was probably confined in the lower atmosphere in closed magnetic structures. The absence of significant coronal mass ejection and II type radio burst rules out the magnetic rope eruption as a mechanism of helioseismic waves.

Helioseismic waves, known as ``sunquakes,'' can be generated by several mechanisms: momentum transfer due to heating of the chromospheric plasma by accelerated energetic charged particles \citep{Kosovichev1995}, impulsive Lorentz force \cite{Fisher2012}, pressure gradient due to flux rope eruption \citep{Zharkov2013}, and rapid dissipation of electric currents \citep{Sharykin2014a}. However, the precise mechanism of sunquakes is still unknown \citep[for a recent review, see ][]{Kosovichev2014}. The main goal of this work is to trace the plasma dynamics and magnetic field changes in the solar atmosphere, as well as possible agents of the flare energy release and sunquake initiation for this particular flare. We use the SDO/HMI data \citep{Scherrer2012} to determine the flare region magnetic field topology, estimate electric currents, and investigate the sunquake. X-ray data from the Ramaty High Energy Solar Spectroscopic Imager (RHESSI) satellite \citep{Lin2002} are used to calculate parameters of accelerated particles, and estimate their impact in the low chromosphere.

\section{GENERAL CHARACTERISTIC OF THE FLARE}

The M9.3 flare on July 30, 2011, occurred in active region NOAA 11261, which had a $\delta$-type magnetic configuration. According to the GOES-15 soft X-ray data (Fig. 1), the flare started at 02:04 UT, reached maximum at 02:09 UT, and ended at about 02:30 UT. The RHESSI observations (Fig. 1) show the presence of the X-ray emission up to 300 keV, reaching maximum at 02:08 UT.

According to the Learmonth and Culgoora radio observatories, there were no type II radio bursts. This means the absence of significant shock waves, which could be associated with erupting magnetic structures. There was no recorded CME that indicates on a non-eruptive nature to the event. The radio data also show the absence of III type radio bursts, thus indicating a closed geometry of magnetic field lines in the flaring region.

The Atmospheric Imaging Assembly (AIA) on SDO \citep{Lemen2012} observes the Sun with time cadence 45\,s spatial resolution $1.2^{\prime\prime}$ (pixel size $0.6^{\prime\prime}$). In Fig. 2 we show the AIA images before and after the flare, illustrating the reconfiguration of ribbon-like structures in the AIA 304 $\rm\AA$ channel ($lgT=4.92$). This reconfiguration probably corresponds to a reduction of the magnetic shear during the flare. In the AIA 94 channel $\rm\AA$ ($lgT=6.86$), we observe the formation of hot large-scale loops. These AIA images are taken at the preflare and postflare times.

The flare signals located in the vicinity of the magnetic field polarity inversion line (PIL) are also detected in all HMI observables (Fig. 3). We will discuss the HMI observations in the next sections.

\section{CONTINUUM EMISSION AND FLOW PATTERN IN THE FLARE REGION}

The flare produced a strong emission observed in the HMI data continuum data. It originated near PIL and expanded in the North-West  direction in a wave-like manner (Fig. 3b). The peak of the HMI continuum intensity averaged over the flare region has a 2-minute delay relative to the HXR maximum (Fig. 1-top). This continuum signal is shown in Fig. 3b by the time-difference images. The estimated velocity of the signal expansion is $\sim 30-40$ km/s, which is higher than the local sound speed and can be comparable with the magneto-acoustic speed of $\sim 20$ km/s.

We also show in Figs. 3b and 3c the line-of-sight velocity structure by contours. The velocity variations are revealed in the Dopplergram differences (Fig. 3a). The most intriguing feature is the strong upflows reaching 3.6 km/s in the vicinity of PIL. These flows are long-lived structures seen before the flare, so they cannot be artifacts due to impulsive variations of the line profile caused by the flare. Besides, we observe sharp changes in the flow structure during the most intensive HXR emission. Such impulsive variations may be associated with the rapid variations of the line profile, and in this case, the corresponding Doppler velocity variations are difficult to estimate. After the flare impulsive phase, the upflow region elongated along PIL is divided into two upflow regions separated by a distance of $\sim 4$ Mm.

\section{MAGNETIC RESTRUCTURING}

The flare is accompanied by significant restructuring of the magnetic field, which resulted in a rapid formation of a small sunspot from an initially diffuse magnetic field. The spot formation happened during the short flare period $<1$ hour (Fig. 4). To our knowledge, this is the first observation of such rapid sunspot formation associated with a solar flare.

Vector magnetograms from HMI also reveal substantial changes of the magnetic field structure in the vicinity of the magnetic neutral line during the flare. In particular, the magnetic shear is sharply reduced (Fig.~5). To illustrate the magnetic field reconfiguration, we calculate a non-linear force-free field extrapolation (NLFFF) of magnetic field using the disambiguated HMI vector magnetograms \citep{Centeno2014} with 720~s time cadence as the initial condition. We use the technique described in \citep{Wheatland2000}, based on minimization of functional $\int_V|\vec{j}\times\vec{B}|^2+|\nabla\cdot\vec{B}|^2dV$. The results of the NLFFF extrapolation for the vector magnetograms before and after the flare are shown in Fig.~5. Initially, a sheared magnetic structure elongated along the PIL (marked by red color) became less sheared after the flare. This corresponds with a transition to a more potential magnetic structure. It appears that these changes in the magnetic topology caused a rapid concentration of the magnetic field and formation of the small sunspot (Fig.~4).

Previous observations showed the disappearance of small sunspot-like structures after some flares \citep[e.g.][]{Wang2002,Chen2007}, but this event is probably the first observation of the formation of such a magnetic structure caused by flare magnetic restructuring.

\section{SUNQUAKE}

The sunquake event was initially revealed as a circular shaped wave in the running difference of the Doppler velocity data. The sunquake wave (Fig.~6)  is best visible about 20 minutes after the initial flare impact on the photosphere located near the PIL (Fig. 3). The wave front was highly anisotropic. The wave front  traveling outside the magnetic region to the South-East had the highest amplitude. In the  North-West direction, the wave traveled towards a big sunspot, and its amplitude was suppressed when it reached the sunspot.

Figure 6b illustrates the positions of the flare impact and the helioseismic fronts observed in the Doppler-shift data, filtered for a central frequency of 6 mHz. The time-distance diagram \citep{Kosovichev1998,Zharkova2007} for the helioseismic wave traveling toward the South-East is shown in Fig.~6a. The dashed curve represents the theoretical time-distance relation calculated in the ray approximation for a standard solar model. The location of this curve is chosen to approximately match the leading wave front. The short strong signal at the beginning of the event indicates the initial source motion.

The time-distance diagrams show two interesting features. First, during approximately the first three minutes, the wave source moved with a speed of about 15-17 km/s, which is higher than the local sound speed but may correspond to the  magneto-acoustic speed in the sunspot penumbra in the vicinity of the source. Second, this motion may be associated with a supersonic plasma eruption, or caused by a series of impacts due to a fast moving magnetic reconnection process. The source motion is a common feature of sunquakes, and may be essential for generating high-amplitude helioseismic waves in solar flares \citep{Kosovichev2014}. The strongest waves are usually observed in the direction of the source motion \citep{Kosovichev2006}.

\section{ELECTRIC CURRENTS IN THE FLARE REGION}

In this section, we consider the evolution of the electric currents at the photosphere level. To estimate the horizontal electric currents, we use Faraday's law applied to the 45-second line-of-sight HMI magnetograms with the spatial resolution $1^{\prime\prime}$ and pixel size $0.5^{\prime\prime}$:

\vskip-.6cm
\begin{eqnarray}
\oint_C\vec{E}\cdot\vec{dl} = -\frac{1}{c}\frac{d}{dt}\left(\int_{S_C}\vec{B}\cdot \vec{dS}\right)
\end{eqnarray}

We can estimate the average transversal component of electric field $\left<E_{\perp}\right>=[d\Phi_z/dt]/cL$, where $\Phi_z$ is the total magnetic flux inside a contour with length $L$, which covers the flare region. The evolution of $d\Phi_z/dt$ presented in Fig.1 (histogram in bottom panel) shows that the whole flare impulse correlates with $d\Phi_z/dt$.

To calculate the vertical currents we use the disambiguated HMI vector magnetic field data \citep{Centeno2014}, and the Ampere's law \citep{Guo2013}:

\vskip-.6cm
\begin{eqnarray}
j_z=\frac{c}{4\pi}(\nabla\times\vec{B})_z = \frac{c}{4\pi}\left(\frac{\partial B_x}{\partial y} - \frac{\partial B_y}{\partial x}\right)
\end{eqnarray}

The resulted $j_z$ maps, effectively averaged over 12 min due to the HMI temporal resolution, are presented in Fig.~7. We can see that the enhancement of electric currents partially correlates with the location of the strongest photospheric impact.

To estimate the measurement error, we calculate the distribution of electric currents in a non-flaring region, and then approximate by a Gaussian. The resulted standard deviation is considered as error of $j_z$ due to noise. The time evolution of $\left<j_z\right>$ averaged over the flare region (Fig. 8) shows that the maximum of $\left<j_z\right>$ is significantly delayed ($\sim$ 30 min) relative to the X-ray flux observed from GOES. Perhaps, the generation and dissipation of electric currents continued after the flare impulsive phase, and might be associated with the magnetic restructuring. It might also be explained by not accounting for the fine structure of electric currents in the HMI data. The problem is we assume that the observed current density is distributed uniformly through HMI pixels, but it could be concentrated in much thinner tubes. Assuming small values of the filling factor at the flare beginning and its subsequent increase due to the current dissipation, one could fit $\left<j_z\right>$ to the X-ray time profile. Resolving this issue requires high-resolution spectro-polarimetric observations.

\section{RHESSI IMAGES AND SPECTRA}

To determine properties of the accelerated particles and the hot flare plasma, we use the RHESSI data in the range of 5-250 keV. RHESSI employs a Fourier technique to reconstruct X-ray emission sources \cite{Hurford2002}. We applied the CLEAN algorithm to synthesize the X-ray images. In Fig. 9 (right panel), the RHESSI HXR and SXR contour images are compared with the corresponding AIA 94 \AA ~images for the time interval covering the HXR peaks of both flares. We choose this AIA channel because it experiences less saturation during the impulsive phase. The AIA brightnings correlate with the HXR emission sources and reveal a loop-like structure with a more intensive West footpoint.

The power-law approximation $f(E)=AE^{-\gamma}$ (where $A$ is a normalization coefficient for energy 10 keV) is considered for the hard X-ray (HXR) nonthermal emission $\gtrsim 20$ keV. To simulate the presence of the low-energy cutoff, we use the broken power law \citep{Holman2003} with fixed photon spectral index $\gamma_0=1.5$  below the break energy ($E_{low}$).

The thermal soft X-ray (SXR) spectrum at $\lesssim 20$ keV is approximated by a single-temperature thermal bremsstrahlung emission law with two parameters: temperature ($T$) and emission measure ($EM$). The RHESSI spectra are fitted by means of the least squares technique, implemented in the OSPEX package with 5 free parameters ($EM$, $T$, $A$, $E_{low}$, and $\gamma$). Figure 9a (top) displays the observed X-ray spectra and the results of the model fitting with $\chi^2=2.37$.

The total flux of accelerated electrons, $F$ [electrons s$^{-1}$] can be estimated following the work of \citet{Syrovatskii1972}:
\vskip-.6cm
\begin{eqnarray}
F(E_{low}<E<E_{high}) = 1.02\times 10^{34}\frac{\delta_1^2}{E_{low}\beta(\delta_1,1/2)}\frac{I_{ph}(E_{low}<E<E_{high})}{[1-(E_{low}/E_{high})^{\delta_1}]}
\end{eqnarray}
where $E_{high}$ is the upper energy cutoff, $\delta =\gamma - 1 $ is the spectral index of accelerated electrons in the HXR emission region, $\beta(x,y)$ is the beta function, and $I_{ph}(E_{low}<E<E_{high})$ photons s$^{-1}$ cm$^{-2}$ is the energy of the integrated photon spectrum in the range shown in the brackets. From the fitting results using this formula, we obtained electron flux $F\approx (1.6\pm 0.4)\times 10^{37}$ electrons s$^{-1}$. The total energy flux of accelerated electrons is $P_{nonth}= 4.6\times 10^{29}$ erg s$^{-1}$. Theoretically, these electrons could contribute to the sunquake initiation. However, the discrepancy between the locations of the sunquake impact and the strongest HXR emission source indicates that the beam-driven origin of the sunquake is unlikely.

\section{DISCUSSION}

The current paradigm of solar flares is that the magnetic energy is released in the solar corona and transported towards the lower atmosphere by heat and energetic particles. For instance, in the hydrodynamic thick-target model \citep[e.g.][]{Kostiuk1975, Fisher1985, Kosovichev1986, Allred2006, RubiodaCosta2014}, high-energy electrons accelerated in the upper corona are injected along magnetic field lines into the atmosphere, generate hard X-ray emission in the loop footpoints, and heat the upper chromosphere to high temperature, producing a high-pressure region. The high-pressure region expands producing upward and downward propagating shocks. The downward shock may reach the photosphere and cause a sunquake.

However, this M-class flare showed a very strong photospheric and helioseismic signal, but had relatively weak high-energy emissions and coronal dynamics. These observations suggest that a substantial part of the magnetic energy is released directly in the low atmosphere. This requires a new mechanism of energy release and transport into the low atmosphere during the flare impulsive phase. One of the possible alternative mechanisms of sunquake and flare initiation can be a rapid dissipation of electric currents or sharp enhancement of Lorentz force \citep{Fisher2012, Sharykin2014a}. The dynamics and spatial structure of electric currents give evidence in favor of such a hypothesis. However, the relationship between the current dynamics and precesses of particle acceleration is unclear. The observations show that the maximum X-ray fluxes occurred before the mean electric current reached maximum.

The observed wave-like continuum emission probably is associated with the sunquake initiation. The direction of propagation of this wave corresponds to the wave observed in the HMI LOS velocity. The estimated velocity is $\approx 30-40$ km/s, which is higher than photospheric magneto-acoustic speed $\sqrt{v_A^2+c_s^2}\sim 20$ km/s for a 500 G magnetic field.

The observed upflows across the magnetic field near the PIL may play a very important role in flare initiation process. These flows could lead to a transversal electric field $E_t=|\vec{v}\times \vec{B}|/c\sim v_zB_t/c$, generating circulating electric currents between the solar lower atmosphere and the corona (Fig. 10). In the conditions of partially ionized magnetized plasma, ambipolar diffusion can lead to efficient dissipation of impulsive electric currents, since coefficient may be several orders higher than the coefficient of Ohmic diffusion \citep[e.g.][]{Khomenko2012}. To clarify such an idea, a detailed modeling of the flare plasma dynamics and heating using a multi-fluid MHD approach is needed.
One of the main peculiarity of the considered flare is a fast formation of the small sunspot. The magnetic field evolution is described by induction equation:

$$
\frac{\partial{\vec{B}}}{\partial t} = \nabla\times(\vec{v}\times\vec{B})+\mu\Delta\vec{B}
$$

where $\mu=c^2/(4\pi\sigma)$ is magnetic diffusivity for plasma with conductivity $\sigma$. In the case of our flare observed sunspot formation can not be connected with diffusion term as it will lead only to sunspot breakup. Probably flows is the main agent of such magnetic field restructuring. Observed by the HMI vertical upflows probably will contribute to the magnetic field intensification as $\partial_x(v_zB_x)+\partial_y(v_zB_y)$. However, in the case of turbulent medium $\vec{v}=\vec{v_0}+\vec{\delta v}$ and $\vec{B}=\vec{B_0}+\vec{\delta B}$ with averages $|\overline{\vec{\delta v}}| = |\overline{\vec{\delta B}}|=0$, time averaged induction equation has form:

$$
\frac{\partial{\vec{B_0}}}{\partial t} = \nabla\times(\vec{v_0}\times\vec{B_0})+\nabla\times\overline{(\vec{\delta v}\times\vec{\delta B})}+\mu_{turb}\Delta\vec{B_0}
$$

where $\mu_{turb}$ is turbulent magnetic diffusivity which is larger than classical one. Thus, we have three competing processes of magnetic field changing in the flare region: convective generation of magnetic field by laminar (1st term) or stochastic (2nd term) flows and magnetic diffusion (3d term). In a lot of flares we usually observe breakup of sunspots \citep{Wang2002,Chen2007}, probably connected with fast diffusion process (magnetic reconnection), and thus, the convective terms are not very important. In our event the flows generating magnetic field can be more important than diffusion and lead to the formation of a small sunspot from initially diffuse sunspot region. To confirm such ideas and observations we need detailed self-consistent numerical MHD simulations.

\section{CONCLUSIONS}

The main results of the work are the following:

\begin{enumerate}
\item The SDO/HMI observations reveal significant changes of the magnetic field topology, leading to a rapid sunspot formation during the solar flare.
\item The HMI continuum emission with velocity of $\sim$ 30-40 km/s shows a fast wave-like expansion. This wave may be related to a shock generated by the energy release in the low atmosphere.
\item The HMI Doppler shift data detect a persistent plasma upflow across the magnetic field near the polarity inversion line with a subsequent break into two upflow regions after the flare impulsive phase. The upflow may play an important role in maintaining the circulating electric currents contributing to the energy release.
\item The location of the strongest  photospheric disturbance leading to the sunquake does not correlate with the HXR emission source. This is not consistent with the hypothesis of the standard flare model. The sunquake generation place is located in the vicinity of the upflow break up, and partially corresponds to the most intense electric currents.
\end{enumerate}

The presented observations show that the flare energy release in the low solar atmosphere not only associated with the precipitation of high-energy particles from the corona. The detected electric currents may also be responsible for generating the flare event. However, the understanding of the mechanism which causes rapid current dissipation requires high-resolution observations and theoretical modeling.

The work was partially supported by RFBR grant 13-02-91165, President's grant MK-3931.2013.2, NASA grant NNX14AB70G, and NJIT grant.



\clearpage

\begin{figure}[t]
\centering
\includegraphics[width=11cm]{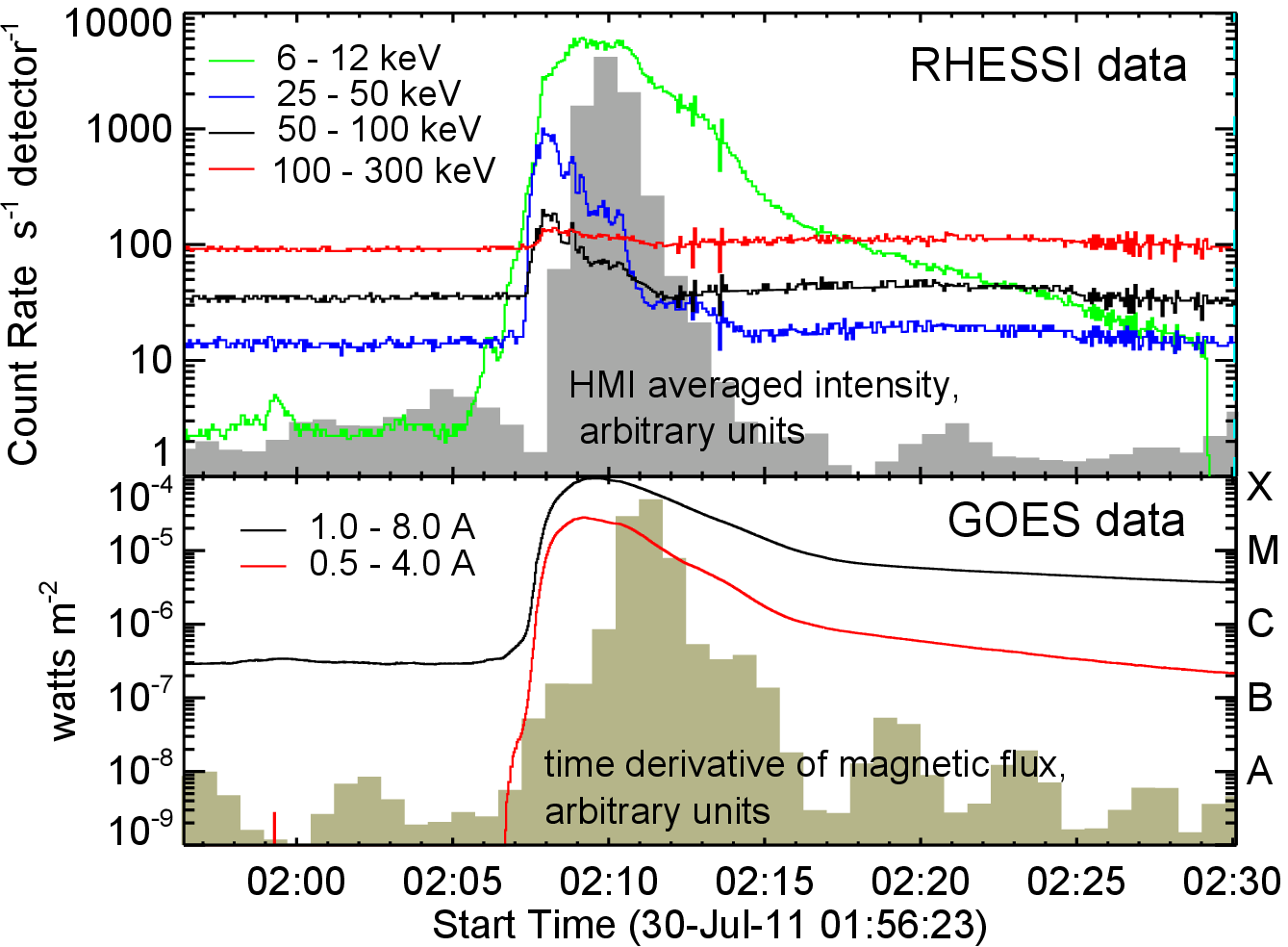}
\caption{Top: color curves show the RHESSI count rates in four different energy bands (marked in the figure); the histogram shows the HMI averaged continuum intensity. Bottom: GOES light curves in two channels 0.5-4 $\rm\AA$ (red) and 1-8 $\rm\AA$ (black); the histogram shows the time derivative of the magnetic flux calculated from the HMI data.}
\end{figure}
\begin{figure}[t]
\centering
\includegraphics[width=0.95\linewidth]{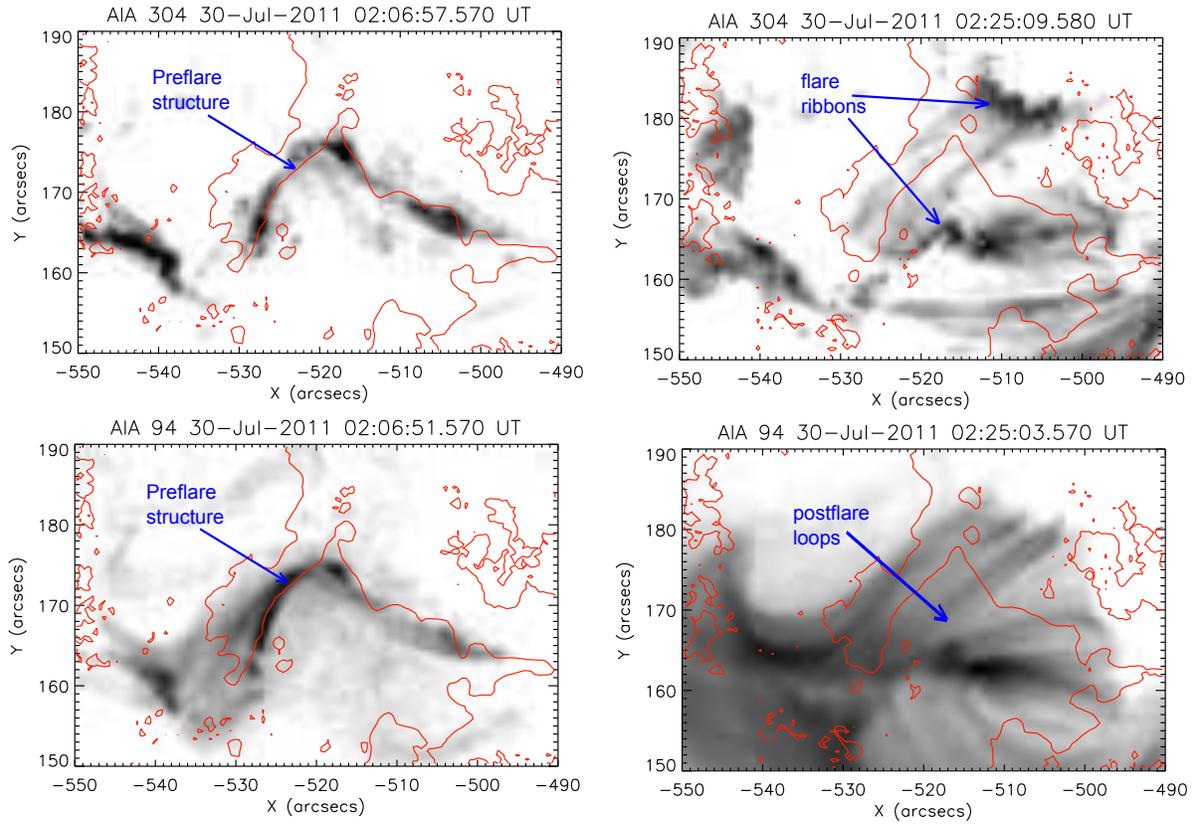}
\caption{The SDO/AIA images in two channels (304 $\rm\AA$, top) and (94 $\rm\AA$, bottom), before (left) and after (right) the flare. Red contour curves show the polarity inversion line.}
\end{figure}
\begin{figure}
\centering
\includegraphics[width=0.7\linewidth]{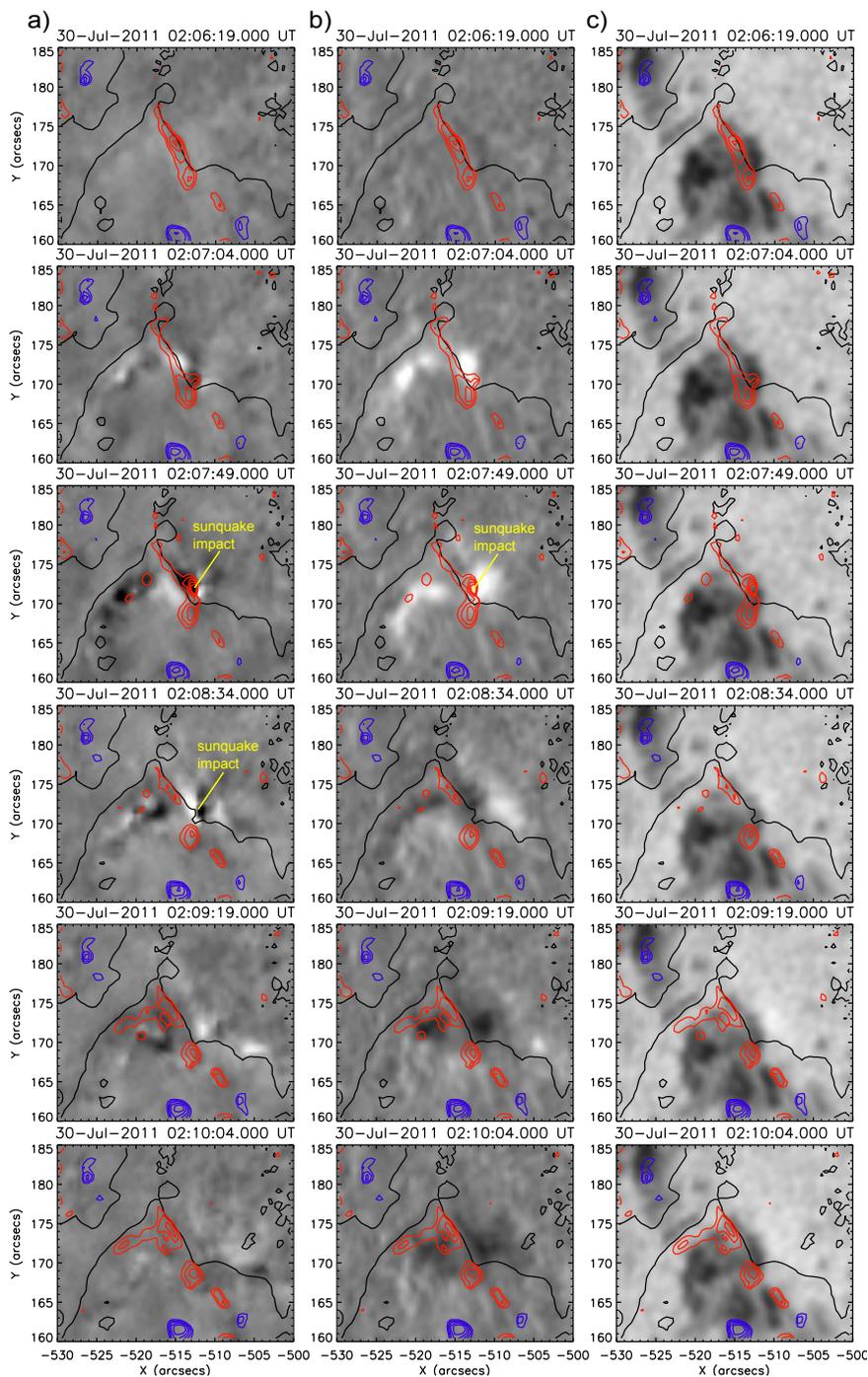}
\caption{HMI images: a) time differences of Dopplergrams, b) time differences of HMI continuum intensity images, and c) sequence of HMI continuum intensity images. Black line shows the polarity inversion line. Red and blue contours show regions of upflows and downflows according to the HMI Doppler measurements.}
\end{figure}
\begin{figure}
\centering
\includegraphics[width=\linewidth]{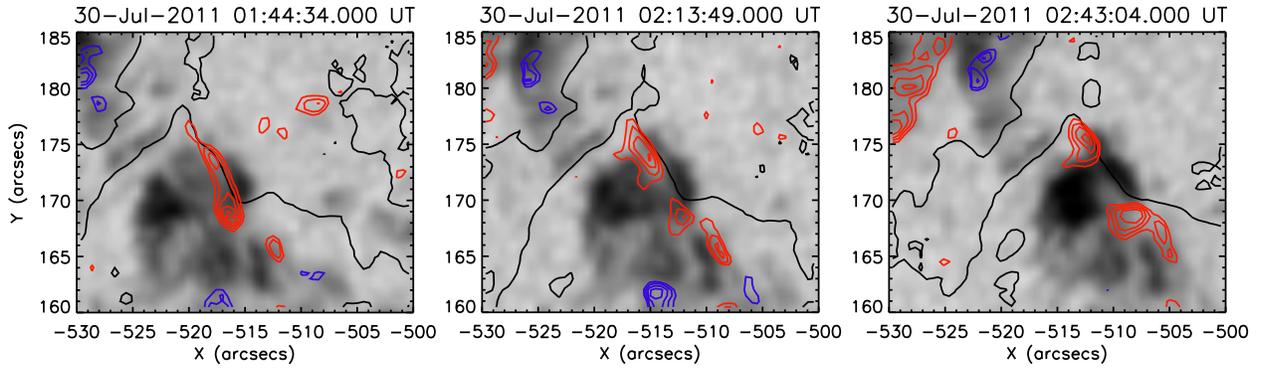}
\caption{Sequence of the HMI continuum intensity images. Black line shows the polarity inversion line. Red and blue contours show regions of upflows and downflows according to HMI Doppler measurements.}
\end{figure}
\begin{figure}[t]
\centering
\includegraphics[width=0.75\linewidth]{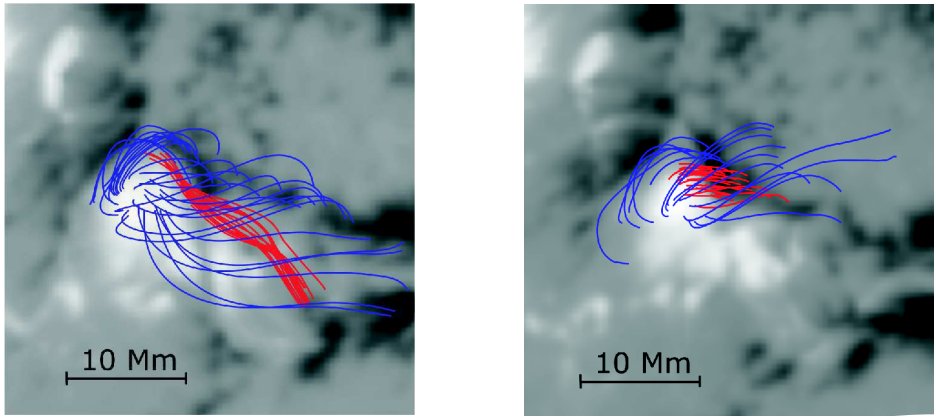}
\caption{NLFFF reconstruction of magnetic field. Left and right panels correspond to preflare and postflare times. Red lines are marked to show the evolution of low-lying magnetic field lines near the magnetic polarity inversion life.}
\end{figure}
\begin{figure}[t]
\centering
\includegraphics[width=0.95\linewidth]{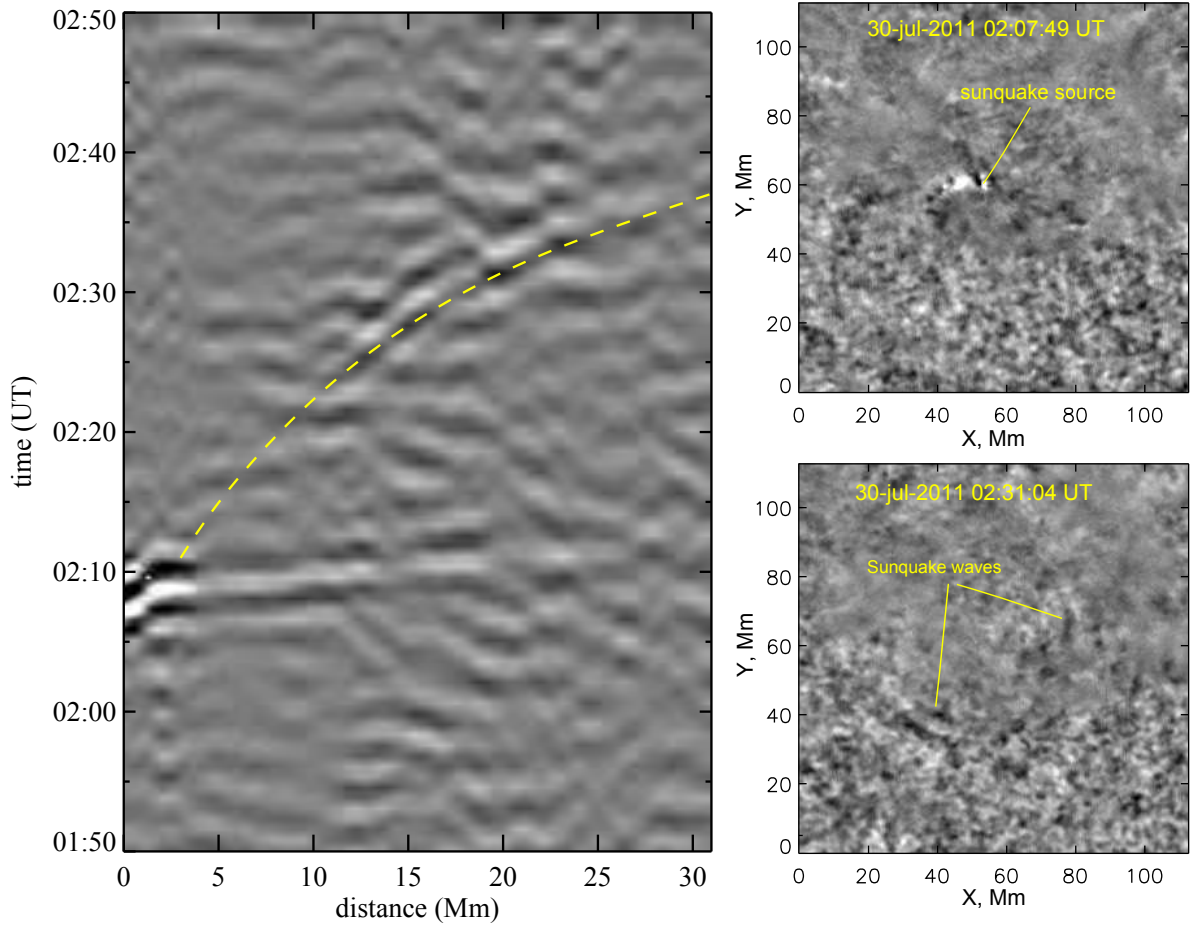}
\caption{ (Left) The sunquake time-distance diagram with the ray-path theoretical prediction (dashed yellow line); (Right) locations of the sunquake source and waves (marked by yellow arrows).}
\end{figure}
\begin{figure}
\centering
\includegraphics[width=\linewidth]{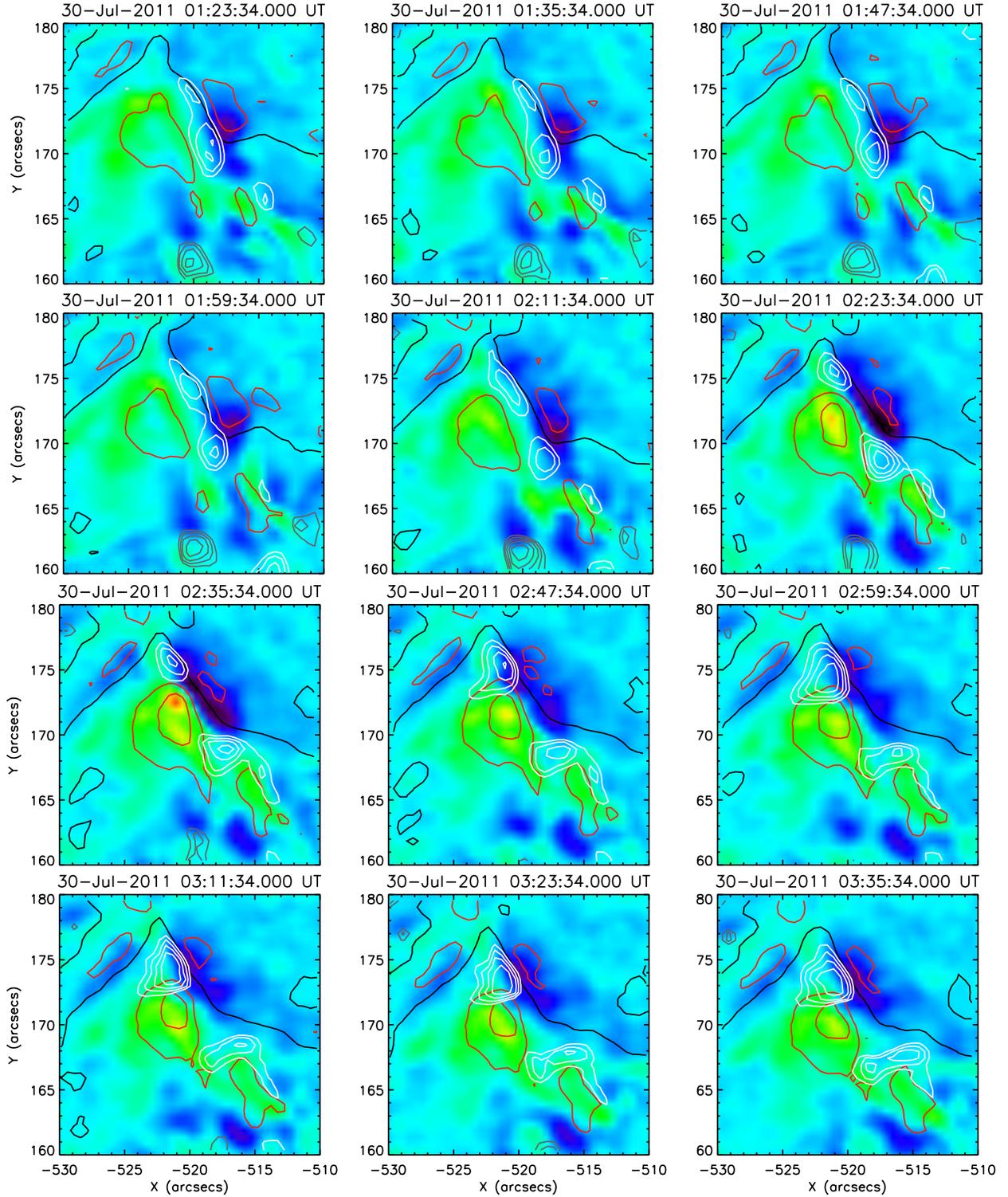}
\caption{Sequence of 12 images which shows the evolution of the vertical component of electric current density (color bar from -0.13 to 0.17 A/m$^2$). Black line shows the polarity inversion line; white and gray contours show upflows and downflows according to the HMI Doppler measurements. Red contours show the LOS magnetic field at 1,000 and 1,500 G levels.}
\end{figure}
\begin{figure}
\centering
\includegraphics[width=0.9\linewidth]{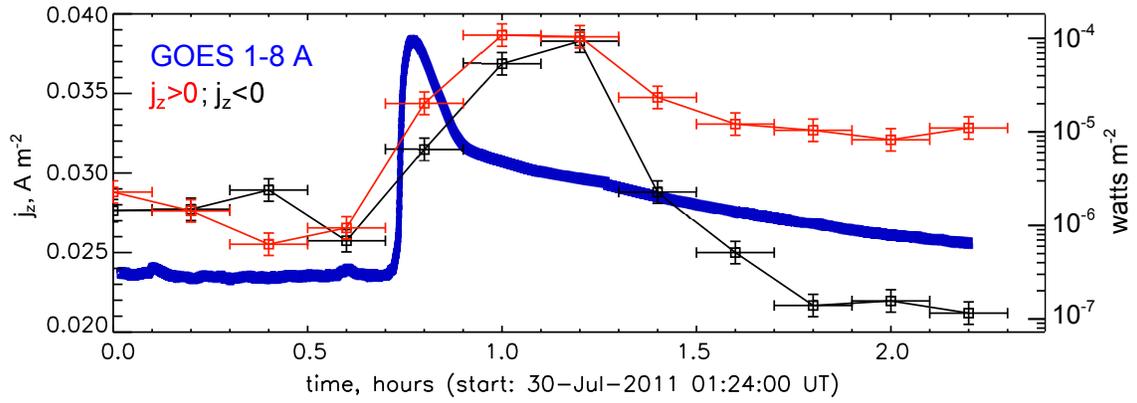}
\caption{Time evolution of the mean vertical current density (red curve shows $\left<j_z\right> >0$, black curve shows $\left<j_z\right> <0$). Blue line is the GOES 1-8 $\rm\AA$ lightcurve.}
\end{figure}
\begin{figure}
\centering
\includegraphics[width=0.9\linewidth]{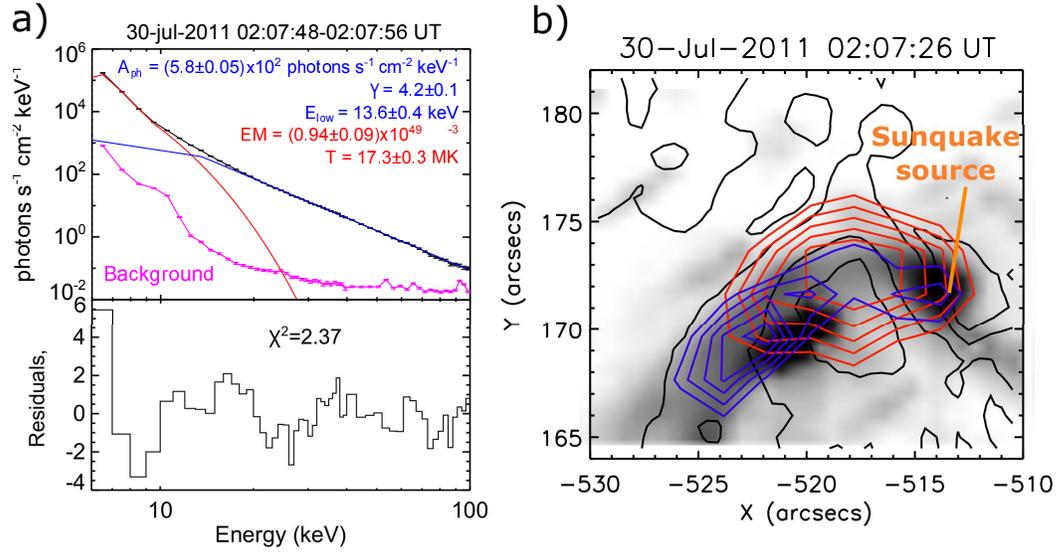}
\caption{a): RHESSI X-ray spectra (upper panel), with the fitting results printed in the plot, the fitting residuals are shown in the bottom panel. b): AIA 94 $\rm\AA$ image with red and blue contours showing the X-ray sources with energies 6-12 and 50-100 keV.}
\end{figure}
\begin{figure}
\centering
\includegraphics[width=0.6\linewidth]{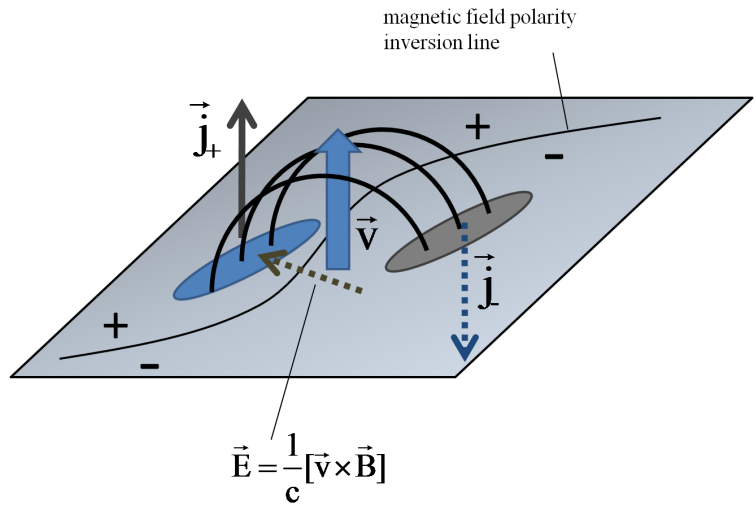}
\caption{The scheme of the flare region near the polarity inversion line.}
\end{figure}
\clearpage
\end{document}